# M-Banking Security – a futuristic improved security approach

Mrs Geeta S. Navale, Mrs Swati S. Joshi , Ms Aaradhana A Deshmukh

Department of Computer Engineering , University of Pune,
Sinhgad Technical Education Society's, Smt. Kashibai Navale  College of Engineering
Pune, Maharashtra State 411041, India

**Abstract**

In last few decades large technology development raised various new needs. Financial sector has also no exception. People are approaching all over the world to fulfill there dreams. Any sector needs to understand changing need of customer. In order to satisfy financial need for customer banks are taking help of new technology such as internet. Only problem remain is of security. The aim of this work is to provide a secure environment in terms of security for transaction by various ways. In order to improve security we are making use of "Steganography" technique in the way never used before. Task of enhancing security include construction of formula for both data encryption and also for hiding pattern. Server should not process any fake request hence concept of custom "Session id" and "Request id" is introduced. Implementation of such a security constraints in banking sector not only help to serve customer in better way but also make customer confident and satisfy.

## 1. Introduction

The importance of an Improving M-Banking Security is both to benefit the Bank as well as the customer. Using Improved M-Banking Security valid user will only access his bank account with the help of mobile phone which has internet facility & the bank application installed. The data exchange between user & bank is carried under protection of encryption & description technique. It makes use of images as cover to hide data with the help of steganography technique.

Here we make use of total 4 different formulae i.e. 2 for steganography, 1 for encryption and 1 for session/request id. Only single key is sent along with image used in steganography (cover image) and used by all 4 formulae in interdependent manner. Key is variable i.e. changed every time and transferred in coded format in steganographed image so that it becomes difficult to get data even if key is found.

## 2. Steganography

The word steganography is derived from the Greek words steganos which means covered and graphie which means writing. Thus, steganography literally means "covered writing." Steganography has been used throughout history for secret communications.

Instead of making use of some previous techniques of steganography, such as LSB method which are easy to detect, it uses two mathematical formulae.  To make use of steganography appropriate image format should be selected such as loss less image formats. There are two types of image formats:

### 2.1. Lossy Image

The image which is to be sent using internet is compressed so the bit pattern in the image will change such a format is called as lossy. This is not desirable in the image as we are hiding the data in image.
Example .GIF, TIFF (.TIF & .TIFF), & BMP (.BMP .DIB) are compressed while sending over network so they can not used right now.

### 2.2. Lossless Image

The images that are not compressed while sending over the network and the bit pattern is not changed are called as lossless image.
These images are suitable for hiding the data in image raster which will not modified while sending over the internet hence we have made use of this images to send the data.
Example: Image files with extension .PNG are lossless.

#### 2.2.1 Cover (Decoy)

Cover serves as medium to hide the message being sent. Making use of graphical images as cover solve many problems readily such as large availability of images, various formats of images and variation in size of image.

Both server and end user application will communicate only by sending the cover image and then extracting data from it.

Instead of making use of some previous techniques of steganography, such as LSB method which are easy to detect, we make use of two mathematical formulae. First formula will generate series of pixel number, with the help of key taken as input, in which data will hide. Second formula will generate the bit number for 'n' bit pixel in which one bit of data will hide.

Only one key is required and need to be sent along with cover image which is used by first formula. While other formulae work in interdependent fashion i.e. result of one serves as input for second.

In order to avoid the detection by comparison of modified image with original image, random pixel other than which contains data are also modified.

### 2.2.2 Implementing steganography

Steganography technique is no more secrete and also by making use various advanced tools, secrete messages in images can also be detected. Some concepts used in steganography makes detection somewhat easy such as sequential use of pixels to hide data, using same bit from 'n' bit pixel to hide data e.g. LSB, availability of original image which is used to compare with modified image etc.

Those loop holes can be avoided by various ways. Making use of mathematical formula is one of them. Here we make use on two mathematical formulae. First formula will generate series of pixel number in which data will hide. Second formula will generate the bit number for 'n' bit pixel in which one bit of data will hide. Using these formulae data will be randomly distributed in image instead of sequential and data hiding bit is also changed every time. Using custom images created by the owner (in this case bank) will help to avoid the comparison of the image.

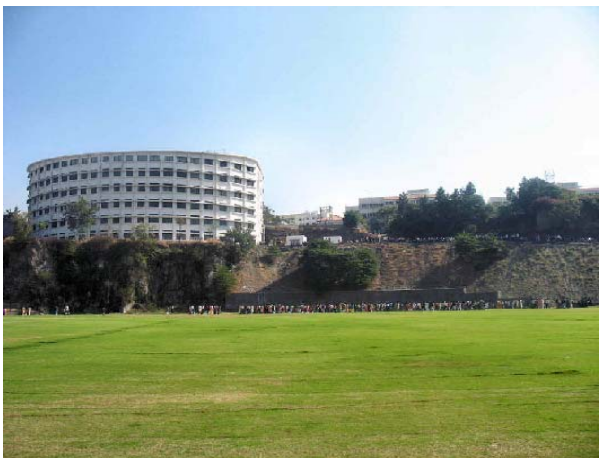
Fig.1 Image before hiding data

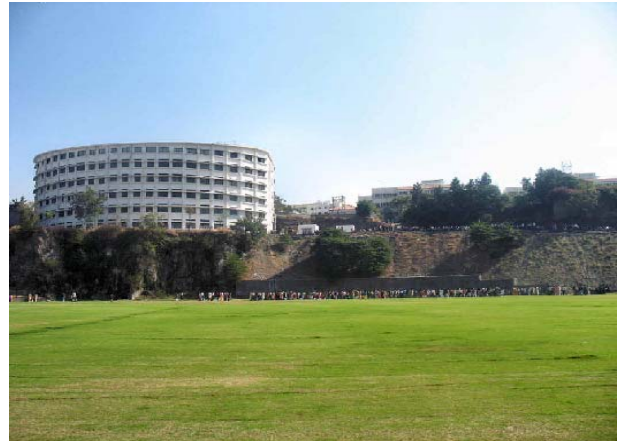
Fig.2 Image after hiding data

## 3. Data Encryption and Error checking

It is necessary to protect the bank transaction even if the images is somehow modified or changed while in network. Concept of encryption and error checking is used to detection of changes in image.

### 3.1 Use of encryption and decryption

Encryption is a process of translating a message, called the Plaintext, into an encoded message. It helps to protect the original data from unauthorized access. Decryption is process of translating encoded message back to plain text. Before embedding the data into image using steganography technique, data is first encrypted either by using simple technique like 0 converted to 1 or any standard algorithm. Such encrypted data is then used for steganography.

### 3.2 Error checking

This concept is somewhat similar to error checking used by network but limited only up to data. Basically if random changes are made in image then resultant extracted data from image may get changed which is undesirable. So error checking in this case means checking consistency of data in image after receiving by receiver.

This can be achieved by carious ways. We can make use of simple techniques such as parity code or any other standard techniques. Data which is required for error checking should also be added to image along with original data.

## 4. Custom session and request id

Now suppose data is encrypted and hidden in image and also successfully transmitted to server. But is it valid request? May be not! Here concept of session and request id introduces. These concepts are being already used in networking while sending and receiving packets. Making some changes in concept helps to improve security for our transaction too.

4.1 Session id

Session id is used to mark or indicate the session is initiated after the authentication while request id is used to make request to process transaction. Before user performs authentication, user application will initiate the connection with server. At that time itself the session id will be issued and it will be temporary. When authentication completes the transaction session id will be initiated. Session id will be constant throughout one session. Server will identify the authorized application's session using session id. Session id will also be derived using formula which uses default key for first transaction.

4.2 Request id

To communicate with server every message hidden in image should have request id. This id will not be sequential fashion instead it will also be generated by formula. We know that key used in this process is variable i.e. changed every time for every message. So the formula for request id will generate id using key for previous transaction.
At server, if session id found valid then request id is checked. If any request with incorrect request id will not be processed. As request id is not sequential next id will not be after equal interval but interval depends indirectly on key. This avoids the fake request to server.

## 5. Data storage capacity of image

Total data to transmit will in 4 part namely session id, request id, actual data and key. Consider the image of 10000 from which 1000 pixels are used. Header will contain session id, request id and key. Header will be in predefined format. Let us consider that header is of 100 bits.
If the character is of 8 bit and 1 bit is stored in each pixel so total 8 pixels required for 1 character.
So 900/8= 112 character in 900 pixels.
So in image of size 300X300 i.e. 90000 pixels and 1/100 of pixels used then minimum 100 characters can be stored with maximum security.

## 6. Algorithm

Including the features mentioned above the resultant algorithmic steps will be as follows:

1. Key is decided for single transaction depending on current status.
2. Using key the series of pixel number in which data will be hidden is derived.
3. Original data in encrypted before being embedded in image.
4. Take raster of image to hide data in pixel. Using steganography technique the encrypted data is embedded in cover image as per pixel locations given in step 2.
5. Request id is generated for transaction depending on key.
6. Finally image is transferred to destination with request id generated in step 5.

Entire working of algorithm is show in fig. 3

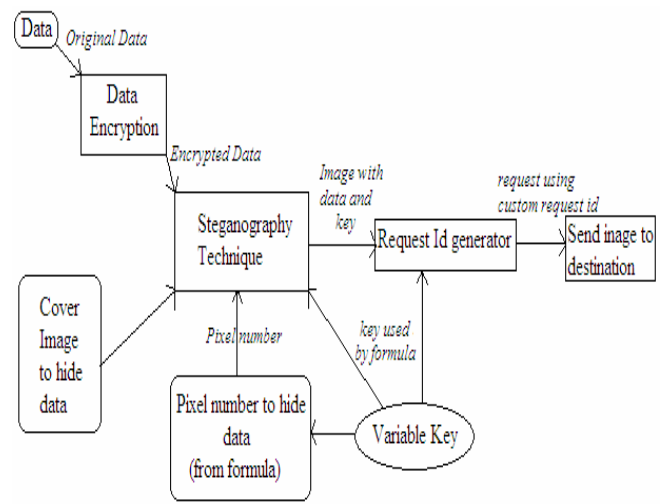

Fig. 3  Working of algorithm

Now before processing request at server side following constraints are checked:

1. First the session id is conformed.
2. If found correct then expected request id and current request id is checked.
3. Request id is correct then key is extracted from image and pixel number series is generated.
4. Data is extracted from image using series in step 3.
5. Data is checked for error if error found retransmission request is sent.
6. If no error then data is decrypted and processed.

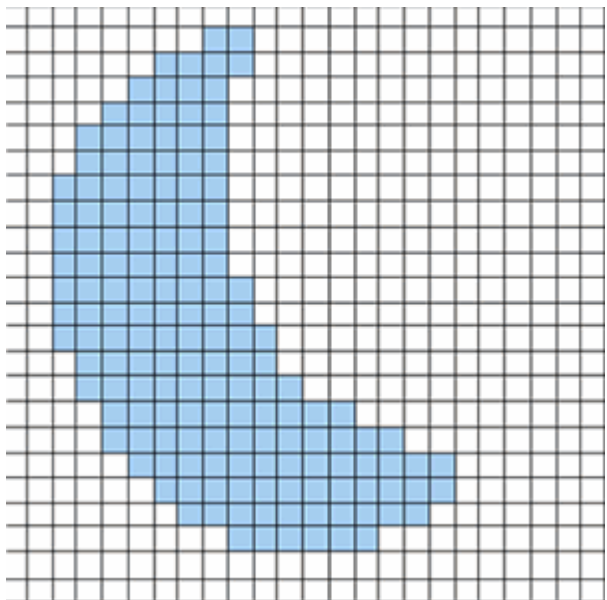

Fig.4 Rater format of image

## 7. Conclusions

Improved M-Banking Security allows user to operate there bank account with the help of mobile like never before. It helps both bank & user to keep there data safe & safe banking. It makes easy to perform various transactions in secured manner. Every year many user face problem related to security of there account and causing loss of valuable information and money too which can be prevented to some extent using combination of various techniques as explained in this short article.